# New results on $\gamma_{\rm str}$ in 2D quantum gravity using Regge calculus *

Christian Holm[a] and Wolfhard Janke[ab] †

[a]Institut für Theoretische Physik, Freie Universität Berlin, 14195 Berlin, Germany

[b]Institut für Physik, Johannes Gutenberg-Universität Mainz, 55099 Mainz, Germany

We study 2D quantum gravity on spherical topologies using the Regge calculus approach. Our goal is to shed new light upon the validity of the Regge approach to quantum gravity, which has recently been questioned in the literature. We incorporate an $R^2$ interaction term and investigate its effect on the value of the string susceptibility exponent $\gamma_{\rm str}$ using two different finite-size scaling Ansätze. Our results suggest severe shortcomings of the methods used so far to determine $\gamma_{\rm str}$ and show a possible cure of the problems. To have better control over the influence of irregular vertices, we choose besides the almost regular triangulation of the sphere as the surface of a cube a random triangulation according to the Voronoi-Delaunay prescription.

## 1. INTRODUCTION

2D Euclidean quantum gravity is believed to be an important toy model on our way to a realistic 4D Minkowskian quantum theory of gravity. Analytic calculations using conformal field theory and matrix models have led to a remarkably good understanding of the 2D theory. In higher dimensions, however, numerical methods, like Regge calculus [1], will probably be indispensable tools to probe quantum gravity. One would therefore like to recover all known continuum results. One interesting aspect is the string susceptibility exponent $\gamma_{\rm str}$, which is the sub-dominant correction to the large area behavior of the partition function $Z(A) \propto A^{\gamma_{\rm str}-3}e^{-\lambda_R A}$, where $\lambda_R$ denotes the renormalized cosmological constant. The exponent $\gamma_{\rm str}$ depends on the genus $g$ of the surface through the KPZ formula $\gamma_{\rm str} = 2 - \frac{5}{2}(1-g)$ [2].

In Regge calculus one introduces a curvature square term and deduces from its expectation value an estimate on $\gamma_{\rm str}$ through a finite-size scaling (FSS) analysis. The model for $R^2$ gravity is defined by

$$Z(A) = \int \frac{\mathcal{D}g}{Vol(Diff)} e^{-S_G} \delta(\int d^2x\sqrt{g} - A), \quad (1)$$

with the gravitational action taken as $S_G = \int d^2x\sqrt{g}(\lambda + \frac{\kappa}{2}R + \frac{a}{4}R^2)$. The coupling constant $a$ sets a length scale of $\sqrt{a}$, and $\hat{A} := A/a$ can be used to distinguish between the cases of weak $R^2$-gravity ($\hat{A} \gg 1$), where the KPZ scaling is recovered, and strong $R^2$-gravity ($\hat{A} \ll 1$) where it was found [3], that $Z(A) \propto A^{\gamma'_{\rm str}-3}e^{-S_c/\hat{A}}e^{-\lambda_R A - b\hat{A}}$, with the classical action $S_c = 16\pi^2(1-g)^2$, some constant $b$, and $\gamma'_{\rm str} = 2 - 2(1-g)$. Note that only for the torus ($g = 1$) weak and strong $R^2$ gravity have the same scaling behavior with $\gamma_{\rm str} = \gamma'_{\rm str} = 2$.

For the sphere ($g = 0$) and the $dl/l$ measure, Gross and Hamber [4] found weak numerical evidence for $\gamma_{\rm str} = -1/2$, while Bock and Vink [5] reported a failure of the KPZ formula. However, in Ref. [6] it was claimed that the string susceptibility exponent $\gamma'_{\rm str}$ for strong $R^2$ gravity was consistent with the theoretical predictions for $g = 0$, but not for $g = 2$. Due to those incompatible results we started to reinvestigate the case of $R^2$ gravity on the sphere in detail.

## 2. METHOD AND FSS ANSÄTZE

We simulated the partition function

$$Z = \int \mathcal{D}\mu(l) \exp\left(-\sum_i (\lambda A_i + a\frac{\delta_i^2}{A_i})\right), \quad (2)$$

where $\mathcal{D}\mu(l)$ is the most commonly used "computer" measure $dl/l$. The notation is identical to that used in Ref. [7]. The deficit angle is given by $\delta_i = 2\pi - \sum_{t \supset i} \theta_i(t)$ with $\theta_i(t)$ being the dihedral angle at vertex $i$, and $A_i$ are barycentric areas. The only dynamical term is the $R^2$-interaction, because we held $A$ fixed during the update. As global lattice topology we used the triangulated surface of a three-dimensional cube [5], and a randomly triangulated sphere constructed according to the Voronoi-Delaunay procedure. In this way we can control the influence of irregular triangulations. For spherical topologies we have the relations $N_0 - 2 = N_2/2$, $N_0 - 2 = N_1/3$, and $2N_1 = 3N_2$, where $N_0, N_1$, and $N_2$ denote the number of sites, links and triangles, respectively.

*Work supported in part by the EEC under contract No. ERBCHRXCT930343 and by the NVV grant bvpf01.
†WJ thanks the DFG for a Heisenberg fellowship.

### 2.1. Finite-size scaling

The methods of Refs. [4,5] have some intrinsic inconsistencies. It was shown in Ref. [5] that the dimensionless expectation value $\hat{R}^2 := a\langle\sum_i \frac{\delta_i^2}{A_i}\rangle$ depends only on $N_2$ and the dimensionless parameter $\hat{A}$. Sending $N_2 \to \infty$ one expects $\hat{R}^2(\hat{A}, \infty)$ to be expandable in a power series whose first three terms read as

$$\hat{R}^2(\hat{A}) = b_0\hat{A} + b_1 + b_2/\hat{A}, \tag{3}$$

where $b_0 = -(\lambda_R - \lambda)a, b_1 = \gamma_{\rm str} - 2$ for $\hat{A} \gg 1$, and $b_0 = -b - (\lambda_R - \lambda)a, b_1 = \gamma'_{\rm str} - 2, b_2 = S_c$ for $\hat{A} \ll 1$.

The scaling Ansatz of Ref. [5] is to consider first a power series expansion of $\hat{R}(\hat{A}, N_2)$ in $N_2$. The expansion of $\hat{R}^2$ is, however, not done at fixed $\hat{A}$, but at a fixed discretization scale set by the dimensionless parameter $\hat{a}_0 := \hat{A}/N_2$:

$$\hat{R}^2(\hat{a}_0, N_2) = N_2 c_0(\hat{a}_0) + c_1(\hat{a}_0) + c_2(\hat{a}_0)/N_2 \ldots \tag{4}$$

A similar good guess would have been to consider an expansion in a linear length parameter $\sqrt{N_2}$. In a second step the coefficients $c_i$ are expanded in $\hat{a}_0$ as $c_0 = c_0^{(0)} + \hat{a}_0 c_0^{(1)} + \ldots$, $c_1 = c_1^{(0)} + \hat{a}_0 c_1^{(1)} + \ldots$, and $c_2 = [c_2^{(0)} + \hat{a}_0 c_2^{(1)} + \ldots]/\hat{a}_0$, and then the continuum limit $\hat{a}_0 \to 0$ is taken. A comparison with (3) then yields $c_0^{(1)} = b_0, c_1^{(0)} = b_1, c_2^{(0)} = b_2$. But because $\hat{a}_0$ is fixed, and not $\hat{A}$, there is no clear control over the crossover from weak $R^2$ scaling behavior ($\hat{A} \gg 1$) to strong $R^2$ scaling behavior ($\hat{A} \ll 1$).

We therefore suggest an alternative approach, where we look at the FSS behavior at a constant value of $\hat{A}$. Expanding $\hat{R}(\hat{A}, N_2)$ at constant $\hat{A}$ we obtain

$$\hat{R}(\hat{A}, N_2) = N_2 d_0(\hat{A}) + d_1(\hat{A}) + d_2(\hat{A})/N_2 + \ldots \tag{5}$$

Also here other FSS Ansätze are possible. The next step is to expand the coefficients $d_i$ as a power series in $\hat{A}$. The coefficient $d_1$ carries all the necessary information to extract the string susceptibilities. A comparison with (3) yields $d_1(\hat{A}) = b_0\hat{A} + \gamma_{\rm str} - 2 + \mathcal{O}(1/\hat{A})$ for $\hat{A} \gg 1$ and $d_1(\hat{A}) = S_c/\hat{A} + \gamma'_{\rm str} - 2 + b_0\hat{A} + \mathcal{O}(\hat{A}^2)$ for $\hat{A} \ll 1$. If we plot $d_1$ versus $\hat{A}$ we expect to see a linear behavior for very large $\hat{A}$, and a divergent behavior for small $\hat{A}$, from which we can extract $\gamma_{\rm str}$ as well as $\gamma'_{\rm str}$.

### 2.2. Simulation parameters

To update the links we used a standard multi-hit Metropolis update with a hit rate ranging from 1 ...3. We ran simulations at constant values of $1/\hat{a}_0$ ranging from 0.2 – 1 280, going very far beyond the range of Ref. [5], where only very small values of $1/\hat{a}_0$ in the range of 0.5 – 5 were studied. The simulations at constant $\hat{A}$ were performed in the range of 14 – 7 800. Usually, the size of the lattices varied from 218 up to 17 498 lattice sites, corresponding to 648 up to 52 488 link degrees of freedom. For each run we recorded about 10 000 measurements of the curvature square $\hat{R}^2$ on every second MC sweep. The statistical errors were computed using standard jack-knife errors on the basis of 20 blocks. The integrated autocorrelation time $\tau_{\hat{R}^2}$ of $\hat{R}^2$ was usually in the range of 5 – 10.

## 3. RESULTS

### 3.1. Scaling at fixed $\hat{a}_0$

For the 4 largest values of $\hat{a}_0$ the scaling Ansatz (4) seems to works well even without the $c_2$ coefficient. A closer look on the curves for $1/\hat{a}_0 = 10$ ($1/\hat{a}_0 = 20$), see Fig. 1, shows apparently two scaling regions, divided approximately by a line through $1/N_2 \approx 0.0005$ (0.0003). We interpret this region as the crossover region from $\hat{A} \gg 1$ to $\hat{A} \ll 1$. Because $\hat{A}$ was chosen to be $N_2/2a$, we decrease $\hat{A}$ either by decreasing $N_2$ or by increasing $a$. This means we always start out on small lattices in the strong $R^2$ gravity regime, and end up on sufficiently large lattices always in the weak $R^2$ gravity regime. It is therefore hard to imagine that one can fit the whole range of data points with the same Ansatz (4). For the lower values of $\hat{a}_0$, the data points of the smaller lattices begin to show a clear deviation from the straight line behavior, which means, that the $c_2$ coefficient, originating from the exponential damping factor of the classical action $S_c$ for the case $\hat{A} \ll 1$, wins more and more. To obtain more data points for small $\hat{A}$ we ran these simulations on lattices as small as $N_0 = 26$. We then fitted all data points using Ansatz (4). In Refs. [5,6] it is claimed that in the limit of $\hat{a}_0 \to 0$, $c_1$ should approach $\gamma_{\rm str} - 2$, and the value of $c_2\hat{a}_0$ should then approach $16\pi^2$. This is, of course, only partially true, because technically this limit is approached by increasing $a$, and therefore decreasing $\hat{A}$, so that strictly speaking $c_1$ should approach $\gamma'_{\rm str} - 2$. Discarding the measurements on the larger lattices leads to an improvement, because they belong to the regime where $\hat{A}$ is large and the $c_2$ coefficient is not so important. One needs very *small* lattices to expect good results. Then, however, one deals with such small lattices, that one needs to worry about finite-size effects.

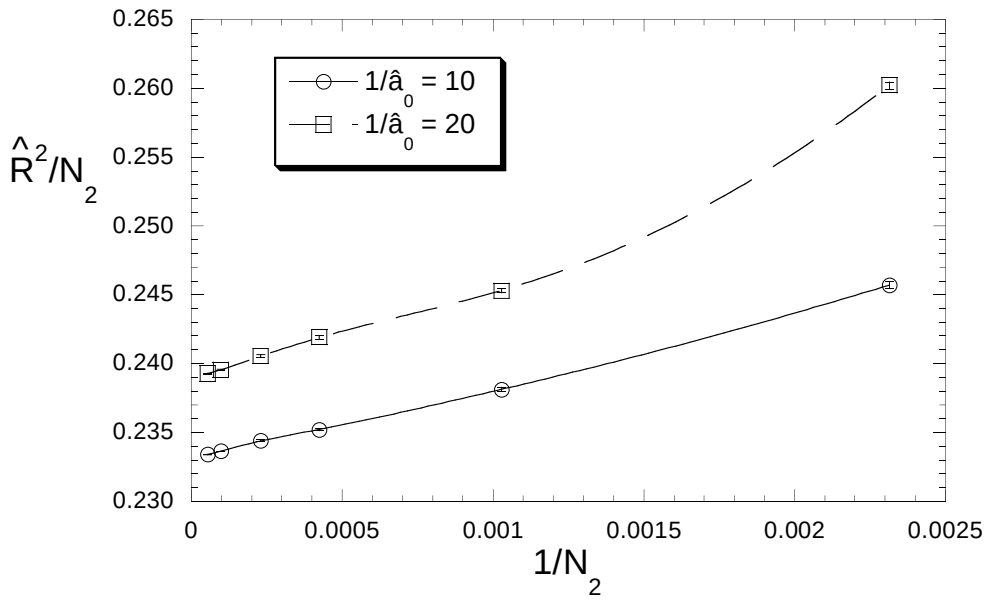

Figure 1. Scaling of $\hat{R}^2/N_2$ versus $1/N_2$ for $1/\hat{a}_0 = 10$ and $1/\hat{a}_0 = 20$.

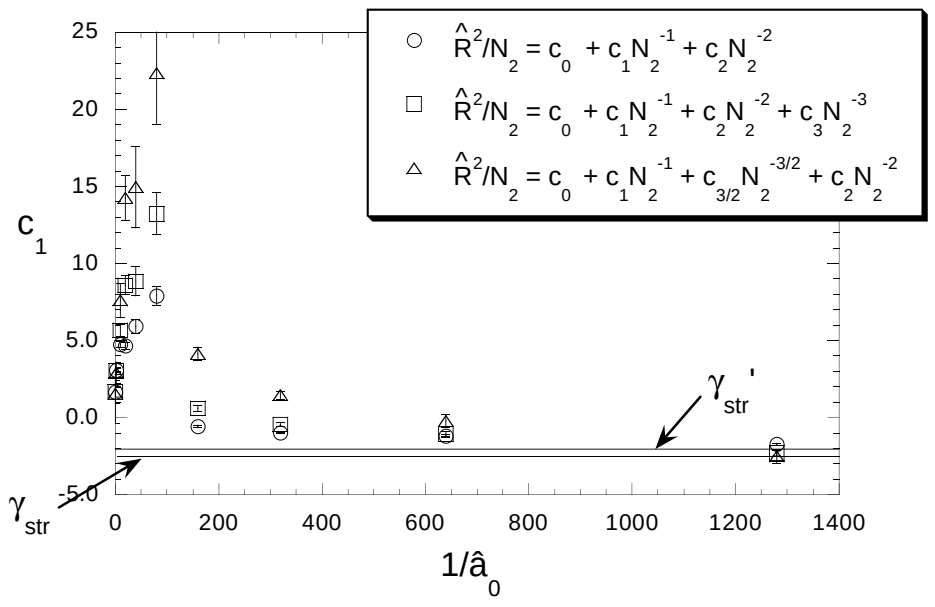

Figure 2. $c_1$ versus $1/\hat{a}_0$ for different FSS Ansätze.

Table 1
Fit results for Ansatz Eq. (4) ($\hat{a}_0 = 1/2a$)

| *dof* | $a$ | $c_0$ | $c_1$ | $c_2\hat{a}_0$ | $\chi^2$ |
|---|---|---|---|---|---|
| 3 | 80 | 0.2530(4) | -2.0(2) | 158.42(6) | 11 |
| 5 | 160 | 0.2520(2) | -1.7(6) | 158.14(2) | 16 |
| 6 | 320 | 0.2522(2) | -1.6(1) | 158.04(2) | 17 |
| 6 | 640 | 0.2535(2) | -2.1(1) | 158.00(1) | 75 |

The values of the fit parameters along with the number of degrees of freedom ($dof$) for an acceptable quality of the fit can be found in Table 1. As a final estimate of this analysis we take the average of the four MC estimates. We feel that $\gamma'_{\rm str} - 2 = -1.9(3)$ is a conservative estimate that is nevertheless still consistent with the prediction $\gamma'_{\rm str} = 0$. Another way to improve the quality of the fits is to include the next order correction term to (4). We added two possible correction terms, one of the form $c_{3/2}(\hat{a}_0)/N_2^{1/2}$ and one of the form $c_3(\hat{a}_0)/N_2^2$. In Fig. 2 one can see the values of the coefficient $c_1$ obtained according to three different FSS Ansätze. For small values of $\hat{a}_0$ all three curves seem to come closer together, and are approximately around the theoretical value of $\gamma'_{\rm str}$. For large values of $\hat{a}_0$ one sees no way how to extract $\gamma_{\rm str}$. If one extrapolates to large values of $\hat{a}_0$, then this means $\hat{A} \gg 1$, but in the same limit the discretization scale becomes very large.

### 3.2. Scaling at fixed $\hat{A}$

Looking at the raw data in the plot $\hat{R}^2/N_2$ versus $1/N_2$, see Fig. 3, we first note that all curves for different values of $\hat{A}$ are significantly curved. Straight line behavior is visible only asymptotically for large values of $N_2$.

To determine the coefficient $d_1$ we used Eq. (5), and also a three-parameter fit of the form

$$\hat{R}(\hat{A}, N_2) = N_2 d_0(\hat{A}) + d_1(\hat{A}) + d_{3/2}(\hat{A})/N_2^{1/2}, \quad (6)$$

which yielded a better, but still fairly large total $\chi^2$. In Fig. 4 we observe that qualitatively both curves fulfill the theoretical expectations, namely they show a divergence at small $\hat{A}$ and a flattening slope at large $\hat{A}$. For large $\hat{A}$ we note that no clear linear slope can be observed, which suggests the presence of FSS corrections to scaling. Unfortunately, there are not enough data points to obtain a precise estimate for either $\gamma_{\rm str}$ or $\gamma'_{\rm str}$.

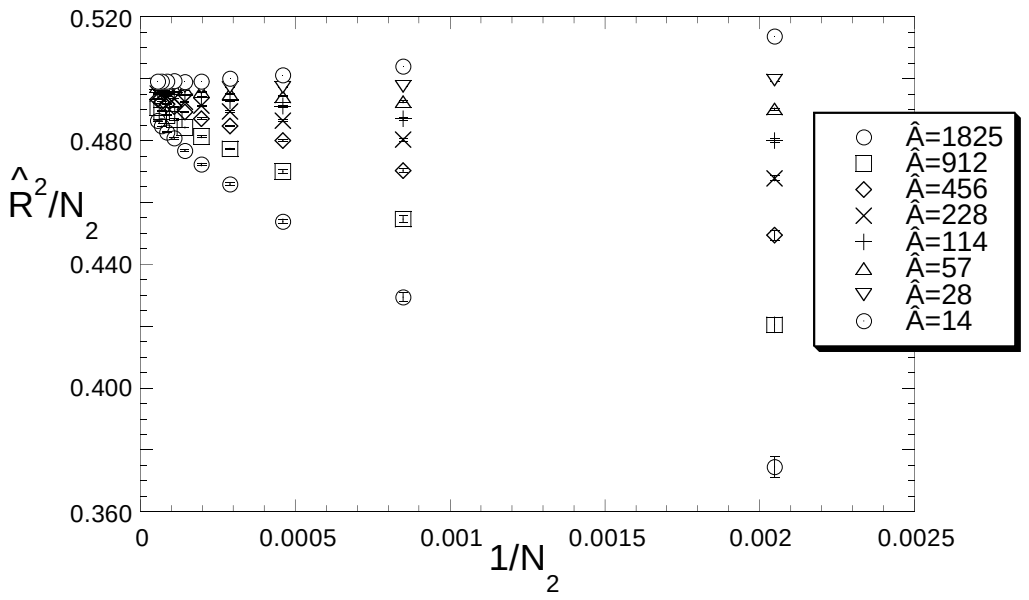

Figure 3. $\hat{R}^2/N_2$ versus $1/N_2$ with $\hat{A}$ = const.

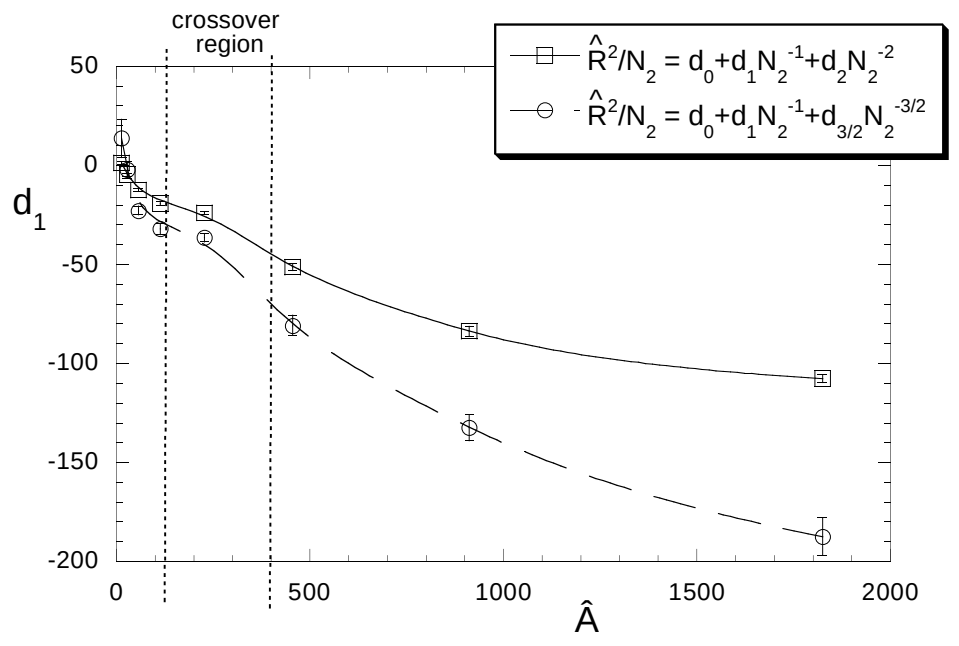

Figure 4. $d_1$ versus $\hat{A}$ for the randomly triangulated sphere for two different FSS Ansätze.

## 4. CONCLUSION

We tried to measure the string susceptibility exponent $\gamma_{\rm str}$ for spherical topologies using two FSS Ansätze in $\hat{R}^2$. Using the irregular triangulated surface of a cube and a randomly triangulated sphere, we observed no qualitative difference between the two types of lattices. The results we obtained for the $dl/l$ measure show that the Ansatz (4) is not applicable to determine $\gamma_{\rm str}$. The extrapolation of $\hat{a}_0 \to 0$ can only predict $\gamma'_{\rm str}$, but with the caveat in mind that already in the first extrapolation (4) one mixes data with small and large $\hat{A}$. Moreover, to reach a sufficiently low value of $\hat{A}$, one has to extrapolate to very small system sizes. In this way one will necessarily encounter large finite-size corrections. As indicated by our result of $\gamma'_{\rm str} = 0.1(3)$, the corrections seem to be unimportant for the sphere. However, we suggest that they are the source of the "failure" for the bi-torus in Ref. [6].

In principle the Ansatz to work at a well controlled $\hat{A}$, as in (5), should be capable to predict both $\gamma_{\rm str}$ and $\gamma'_{\rm str}$, but we have experienced large FSS corrections to scaling, which make it very difficult to extract the coefficient $d_1$ with high accuracy. The results we obtained can still be consistent with the theoretical prediction of Ref. [3] for both, $\gamma_{\rm str}$ and $\gamma'_{\rm str}$.

In light of these difficulties it seems to us very questionable if the method of using FSS properties of the expectation value of the $R^2$ term will ever reach a high enough accuracy to probe the formulas for $\gamma_{\rm str}$ and $\gamma'_{\rm str}$. One needs to develop more direct approaches to measure $\gamma_{\rm str}$, as has been done for the DTRS method [8]. The main conclusion is that for the $dl/l$ measure a failure of Regge calculus has not yet been shown. Details of this study will be presented elsewhere [9].

## REFERENCES

1. T. Regge, Nuovo Cimento 19 (1961) 558.
2. V.G Knizhnik, A.M. Polyakov, and A.B. Zamalodchikov, Mod. Phys. Lett. A3 (1988) 819.
3. H. Kawai and R. Nakayama, Phys. Lett. B306 (1993) 224.
4. M. Gross and H. Hamber, Nucl. Phys. B364 (1991) 703.
5. W. Bock and J. Vink, Nucl. Phys. B438 (1995) 320.
6. W. Bock, Nucl. Phys. B (Proc. Suppl.) 42 (1995) 713.
7. C. Holm and W. Janke, Phys. Lett. B335 (1994) 143.
8. J. Ambjørn, in proceedings of the 1994 Les Houches Summer School, Session LXII, and references therein.
9. C. Holm and W. Janke, work in progress.